\documentstyle[11pt,newpasp,twoside,epsf]{article}

\def\edcomment#1{\iffalse\marginpar{\raggedright\sl#1\/}\else\relax\fi}
\reversemarginpar

\begin{document}
\title{Carbon-rich, extremely metal-poor Population II stars}
\author{Sean G. Ryan}
\affil{Department of Physics and Astronomy, The Open University, Walton Hall, Milton Keynes, MK7 6AA, United Kingdom; s.g.ryan@open.ac.uk}

\begin{abstract}

A significant fraction of very metal-poor stars have unexpectedly high
carbon abundances. I recap on their discovery, describe the
results of high-resolution spectroscopic studies of their composition,
and consider
possible origins. It is now clear that at least two, and possibly three
different explanations are needed for their general characteristics.
\end{abstract}

\section{Introduction}

Beers, Preston \& Shectman (1985,1992) conducted an objective-prism survey 
to search for low-metallicity
stars. The spectral range, defined by an interference filter,
spanned 150~\AA\ near  3950~\AA. This included the
Ca H and K lines (3933~\AA, 3968~\AA), the latter blended with
H$\epsilon$, but did not extend to 
the CH bands around 4310 and 4324~\AA, and had no significant
sensitivity to the CN~band at 3883~\AA.
From these spectra,
Beers et al. selected a weak-lined stars for further 
study. 

Subsequent slit spectroscopy at 1~\AA\ resolution 
identified an unusually high proportion of objects
having very strong CH bands. In a sample of 1044 stars, 
Beers et al. (1992) identified 50 with anomalously strong
CH bands ``characteristic of the subgiant CH stars discussed by Bond''. 
They noted that the proportion of C-rich objects increased at
lower metallicity, being 6 out of 70 at [Fe/H] $<$ $-3$.
From that sampling, it appeared than 10\%\ of their most metal-poor
stars had substantial carbon excesses. Other objects
in their sample might also have carbon excesses but at more moderate levels. 

\section{Element abundances in C-rich stars}

High-resolution studies of C-rich stars in the Beers et al. survey
(and its successors) have found diverse chemical properties.
Sneden et al. (1994) identified a strong excess of r-process elements
in the C-rich star CS~22892-052. 
In contrast, Norris, Ryan \&  Beers (1997a) and 
Barbuy et al. (1997)
identified stars with
carbon-to-iron ratios up to 100 times the solar value in objects with
[Fe/H] $\simeq$ $-2.7$, but with strong s-process signatures. 
This supported
the suggestion by Beers et al. that their C-rich 
objects were related to Bond's (1974)
CH subgiants, which are believed to be 
contaminated by C- and s-process-rich material from a now-extinct
AGB companion. The r-process enhancement in CS~22892-052,
far from being typical, now appears anomalous. 

Is an s-process enhancement the norm? Apparently not.
C-rich stars without anomalous abundances amongst the
neutron-capture elements have also been found (e.g. Norris et al. 1997b).
So have stars with only mild C-enhancements and no neutron capture
anomalies. The situation is summarised by Aoki et al. (2002, their 
Figure~8). Clearly, different processes are required to explain the
s-process-rich and s-process-normal objects, and yet
another process is required to explain the r-process-rich object, CS~22892-052.
Three other r-enhanced objects are known, but they are not C-rich.
The fractions of r-enhanced objects in the C-rich and
C-normal classes are difficult to determine reliably with such small
statistics, but there is no evidence that these fractions differ
greatly.  Aoki et al. (2002) have proposed that the r-process
enhancement in CS~22892-052 may have no causal link to its C enhancement.

\section{Degrees of carbon enrichment}

Published evidence concerning the degree
of C enhancements must be drawn either from the highly-selected objects
for which detailed abundance calculations have been made (citations above),
or from the
low-resolution results of Rossi, Beers \& Sneden (1999, their Figure~2).
The data of Rossi et al. are in agreement with the levels of enhancement
from detailed measurements, but cover a wider range of 
metallicities. Curiously, the data are suggestive of an
upper envelope to the C enrichment, corresponding to a fixed C/H ratio
at [C/H] $\simeq$ $-1$. If such a maximum does exist, it may begin to
explain why the C-rich stars become more noticeable at lower metallicity,
since there the abundance of other metals becomes insignificant in
comparison.

Christlieb et al. (2002) have identified a star, HE0107-5240, 
with an unprecedented iron deficiency, [Fe/H] = $-5.3\pm0.2$.
More curiously, this object has an extremely high C abundance, quite
out of keeping with the normal ratios for its other elements.
Nevertheless, its [C/H] value is $-1.3\pm0.3$, 
which places it close to the extrapolation
of the envelope described above (see Figure~1).

\begin{figure}
\plotfiddle{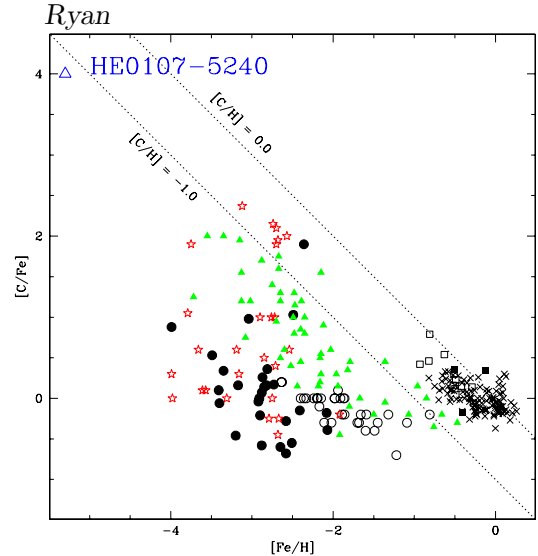}{58mm}{0}{38}{38}{-140}{-070}
\caption{[C/Fe] ratios for a range of stars from the literature.}
\end{figure}

\section{Orbital characteristics}

One prediction of the CH subgiant explanation for the C-rich, s-rich stars
is that the presence of the extinct AGB companion should be detectable 
via radial velocity variations.
The orbital periods of such systems had to be short enough, and hence the 
stars' separations small enough,
to facilitate
accretion of material onto the lower-mass star. However, the stars
must also have sufficiently large separations that they avoided transferring
matter during the donor's first-ascent of the giant-branch,
rather than surviving until the AGB when it attained
an even greater radius (Wallerstein et al. 1997).
These trade-offs suggest periods of 1-10~yr (Han et al. 1995; 
Jorissen \& Boffin 1992). 

In some cases, radial velocity variations on this timescale have been observed.
For example, LP~625-44 has an orbital period of order 13~yr (Tsangarides, Ryan
\& Beers 2003). In contrast, the chemically-similar star
LP~706-7 has revealed
no radial velocity variations over a ten year period. This object
is on the early subgiant branch, cooling from the main-sequence turnoff.
Other main-sequence turnoff stars that are C-rich have also
been found not to exhibit radial velocity variations (Preston \& Sneden 2001).
It is possible that the AGB process is not responsible for the enrichment of
these stars, but if that is so
then the high s-process enhancement in LP~706-7 becomes difficult to explain.

However, it may be that some CH systems become unbound following mass-loss.
Porto de Mello \& Da Silva (1997) have identified the white-dwarf
companion of the Pop.~I Ba dwarf HR~6094 at a separation of 5360~AU.
If such a system can be considered bound, which
is doubtful given its escape speed of $\sim$1~km~s$^{-1}$, then
assuming masses of 1.0 and 0.6~M$_\odot$ implies an orbital period 
3.1$\times$10$^5$~yr. The radial velocity of this Ba star will not vary
on the timescale over which a telescope
allocation committee retains interest.
Perhaps it is possible that some of the
apparently-single, CH-subgiant-like stars have emerged from multiple
systems that also became unbound.

\section{Distinguishing s-rich and s-normal C-enhanced stars}

Aoki et al. (2002, their Figure~8) have examined differences between C-rich 
stars on the basis of their enhancements of s-process elements.
These stars can be further distinguished on the basis of nitrogen enhancements
and carbon isotopic composition (Ryan et al. 2003). In
the samples observed so far, the stars which have normal s-process
abundances are found only high up the first ascent giant branch, whereas s-rich 
stars are distributed over a much wider range of giant and subgiant stages of 
evolution. The s-normal stars also have higher nitrogen abundances.
These features can be reconciled with two distinct origins for the stars,
the s-rich objects being the ones which exhibit material transferred
there by a companion, whereas the s-normal stars presumably began their
lives with high $^{12}$C fractions. These stars have converted some of their 
initial $^{12}$C to $^{13}$C and 
$^{14}$N as they have evolved. In this scenario, the s-rich stars 
are s-rich and C-rich only in their outer layers, whereas the s-normal stars
may be considered C-rich throughout. One consequence of this proposal is
that the s-rich stars will gradually dilute their surfaces as they ascend
the giant branch and hence will become less distinctive, whereas s-normal 
stars will maintain high excesses of the CNO group, albeit with a 
changing isotopic mix. S-normal stars will be found with a 
luminosity distribution that reflects the luminosity selection effects in the
magnitude-limited objective-prism survey. On the other hand, s-rich stars 
will drop out
of the survey as they become more convective, and hence their luminosity
distribution will contain a lower proportion of evolved stars.
This is in the same sense as the difference observed.

\section{Acknowledgments}

I am indebted to my co-workers with whom I have
studied these stars: W. Aoki, H. Ando,
T. C. Beers, R. Gallino, J. E. Norris, and S. Tsangarides. This work
was supported by the allocation committees and staff
of the AAT, WHT, and Subaru,
and PPARC grants PPA/O/S/1998/00658 and PPA/G/O/2001/00041.

\end{document}